\documentclass[showpacs,preprintnumbers,amsmath,amssymb,twocolumn]{revtex4}
\usepackage{graphicx}
\usepackage{dcolumn}
\usepackage{bm}
\usepackage{amssymb}
\usepackage{array}
\usepackage{hhline}
\usepackage{longtable}

\begin{document}

\title{Exact solution for mean first-passage time on a pseudofractal scale-free web}

\author{Zhongzhi Zhang$^{1,2}$}
\email{zhangzz@fudan.edu.cn}

\author{Yi Qi$^{1,2}$}

\author{Shuigeng Zhou$^{1,2}$}
\email{sgzhou@fudan.edu.cn}

\author{Wenlei Xie$^{1,2}$}

\author{Jihong Guan$^{3}$}
\email{jhguan@tongji.edu.cn}

\affiliation {$^{1}$School of Computer Science, Fudan University,
Shanghai 200433, China}

\affiliation {$^{2}$Shanghai Key Lab of Intelligent Information
Processing, Fudan University, Shanghai 200433, China}

\affiliation{$^{3}$Department of Computer Science and Technology,
Tongji University, 4800 Cao'an Road, Shanghai 201804, China}

\begin{abstract}
The explicit determinations of the mean first-passage time (MFPT)
for trapping problem are limited to some simple structure, e.g.,
regular lattices and regular geometrical fractals, and determining
MFPT for random walks on other media, especially complex real
networks, is a theoretical challenge. In this paper, we investigate
a simple random walk on the the pseudofractal scale-free web (PSFW)
with a perfect trap located at a node with the highest degree, which
simultaneously exhibits the remarkable scale-free and small-world
properties observed in real networks. We obtain the exact solution
for the MFPT that is calculated through the recurrence relations
derived from the structure of PSFW. The rigorous solution exhibits
that the MFPT approximately increases as a power-law function of the
number of nodes, with the exponent less than 1. We confirm the
closed-form solution by direct numerical calculations. We show that
the structure of PSFW can improve the efficiency of transport by
diffusion, compared with some other structure, such as regular
lattices, Sierpinski fractals, and T-graph. The analytical method
can be applied to other deterministic networks, making the accurate
computation of MFPT possible.
\end{abstract}



\date{\today}
\maketitle

\section{Introduction}

The problem of random walks (diffusion) is a major theme and
understanding its behavior is central to a wide range of
applications~\cite{HaBe87,MeKl00,MeKl04,BuCa05}. Among various
interesting issues of random walks, trapping plays a significant
role in an increasing number of disciplines, e.g.,
physics~\cite{SoMaBl97,No77}, society~\cite{LlMa01},
computer~\cite{FoPiReSa07}, and biology~\cite{CoTeVoBeKl08}, to name
a few. The classic trapping issue first introduced in~\cite{Mo69} is
a random-walk problem, in which a trap is placed at a given
location, absorbing all walkers that visit it.

An important quantity related to trapping problem is the trapping
time (first-passage time, survival time, or the mean walk length) or
the mean time to absorption. The trapping time of a given site $s$
is the expected time for a walker starting from $s$ to first reach
the trap. This quantity is useful in the study of transport-limited
reactions~\cite{YuLi02,LoBeMoVo08}, target
search~\cite{BeCoMoSuVo05,Sh06} and other physical problems. The
average trapping time, also known as the mean first-passage time
(MFPT), characterizes the process of trapping. It is defined as the
average of survival times over all starting sites.

In the past few decades, there has been considerable interest in
computing the mean first-passage time, in order to obtain the
dependence of this primary quantity on the system size or other
parameters. In a seminal work, using the approach based on
generating functions, Montroll derived the rigorous results for MFPT
of random walks on regular lattices with a variety of
dimensions~\cite{Mo69}. Recently, by applying a decimation
procedure, Kozak and Balakrishnan obtained the accurate solutions
for MFPT on a family of Sierpinski
fractals~\cite{KaBa02PRE,KaBa02IJBC}; using an analogous but
different method, Agliari got the exact expression for the MFPT for
a random walker on T-fractal~\cite{Ag08}. In spite of these rigorous
results, the explicit determination of MFPT for random walks with a
trap on other media is still open~\cite{CoBeMo05}.

It is well established that the scaling of MFPT is related to the
underlying structural properties of the media in which the walkers
are confined~\cite{HaBe87,MeKl00,MeKl04,BuCa05}. Extensive empirical
studies~\cite{AlBa02,DoMe02} have revealed that most real networked
systems share some striking features, such as scale-free
behavior~\cite{BaAl99} and small-world effects~\cite{WaSt98}. These
newly-found properties have a profound effect on almost all
dynamical processes taking place on the
networks~\cite{Ne03,BoLaMoChHw06,DoGoMe08}, including disease
spreading~\cite{PaVe01}, games~\cite{SzGa07},
synchronization~\cite{ArDiKuMoZh08}, and so on. Very recently, a lot
of activities have been devoted to the study of influences of
scale-free and small-world characteristics on the behavior of random
walks, uncovering many unusual and exotic phenomena about random
walks~\cite{PaAm04,NoRi04,SoRebe05,Bobe05,CoBeTeVoKl07,GaSoHaMa07,BaCaPa08,KiCaHaAr08,CaAb08}.
However, to best of our knowledge, rigorous solution for MFPT of
trapping problem on scale-free small-world networks is missing.

In this paper, we study the classic trapping problem on a
deterministic network, called pseudofractal scale-free web
(PSFW)~\cite{DoGoMe02}. We focus on a peculiar case with the trap
fixed at a hub node (node with the highest degree). The PSFW is a
very useful toy model that captures simultaneously scale-free
small-world properties, thus provides a good facility to investigate
analytically trapping process upon it. We derive an exact formula
for the mean first-passage time characterizing the trapping process.
The analytic approach is based on an algebraic iterative procedure
obtained from the particular construction of PSFW. The obtained
rigorous result shows that the MFPT grows as a power-law function of
the number of network nodes with the exponent less than 1, which
implies that in contrast to regular lattices, Sierpinski fractals,
and T-graph, the PSFW tends to speed up the diffusion process. Our
study opens the way to theoretically investigate the MFPT of a
random walker on a wide range of deterministic
networks~\cite{BaRaVi01,RaBa03,ZhZhChGu07,ZhZhFaGuZh07}.

\section{The pseudofractal scale-free web}

Here we introduce the pseudofractal scale-free web defined in a
recursive way~\cite{DoGoMe02}, which has attracted an amount of
attention~\cite{CoFeRa04,ZhRoZh07,RoHaAv07,ZhZhCh07}. We investigate
the PSFW model because of its intrinsic interest and its
deterministic construction, which allows one to study analytically
their topological properties and dynamical processes on it. 

\begin{figure}
\begin{center}
\includegraphics[width=.90\linewidth,trim=60 15 60 0]{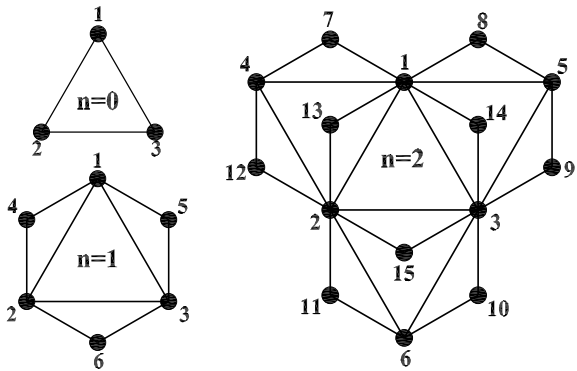} \
\end{center}
\caption[kurzform]{The first three generations of the pseudofractal
scale-free web.} \label{network}
\end{figure}

The pseudofractal scale-free web, denoted by $G_{n}$ after $n$
($n\geq 0$) generation evolution, is constructed as follows. For
$n=0$, $G_{0}$ is a triangle of edges connecting three nodes
(vertices, sites). For $n\geq 1$, $G_{n}$ is obtained from
$G_{n-1}$: every existing edge in $G_{n-1}$ introduces a new node
connected to both ends of the edge. Figure~\ref{network} illustrates
the construction process for the first three generations.

According to the network construction, one can see that at each step
$n_i$ ($n_i\geq 1$) the number of newly introduced nodes is
$L(n_i)=3^{n_i}$. From this result, we can easily compute the
network order (i.e., the total number of nodes) $V_n$ at step $n$:
\begin{equation}\label{Nt}
V_n=\sum_{n_i=0}^{n}L(n_i)=\frac{3^{n+1} +3}{2}.
\end{equation}

Let $k_i(n)$ be the degree of a node $i$ at time $n$, which entered
the network at step $n_i$ ($n_i\geq 0$). Then
\begin{equation}\label{ki}
k_i(n)=2^{n-n_{i}+1}.
\end{equation}
From Eq.~(\ref{ki}), one can easily see that at each step the degree
of a node doubles, i.e.,
\begin{equation}\label{ki2}
k_i(n)=2\,k_i(n-1).
\end{equation}

The PSFW presents some typical characteristics of real-life networks
in nature and society. It has a power-law degree distribution $P(k)
\sim k^{-\gamma}$ with the exponent $\gamma=1+\frac{\ln 3}{\ln
2}$~\cite{DoGoMe02}. Its average path length (APL), defined as the
mean of shortest distance between all pairs of nodes, increases
logarithmically with network order~\cite{DoGoMe02,ZhZhCh07}. In the
large network order limit, the average clustering coefficient tends
to $\frac{4}{5}$. Thus, the PSFW exhibits small-world
behavior~\cite{WaSt98}. In addition, its node degree correlations
are negative, the average degree of nearest neighbors, $k_{\rm
nn}(k)$, for nodes with degree $k$ is approximately a power-law
function of $k$ with a negative exponent $-(2-\frac{\ln 3}{\ln
2})$~\cite{ZhRoZh07}.

While the graph introduced above visually looks very similar to a
fractal (see Fig.~\ref{network}), the similarities are only
superficial. The distinction is obvious. Fractals have a finite
dimension (i.e., their APL grows as a power of the vertex
number)~\cite{Ma82}. In sharp contrast, the network under
consideration has an infinite dimension (i.e., its APL increases
slower than any power of the network order)~\cite{SoHaMa05}. This is
why it is called {\em pseudofractal scale-free web}.

After introducing the PSFW, in what follows we will study the mean
first-passage time for random walks with a single immobile trap on
the web, by applying a method similar to but different from that
introduced in~\cite{KaBa02PRE,KaBa02IJBC}.

\section{Formulation of the problem}

In this section we formulate the problem of a simple unbiased
Markovian random walk of a particle on PSFW $G_n$ in the presence of
a trap or a perfect absorber located on a given node. In order to
distinguish different nodes, we label all the nodes belonging to
$G_n$ in the following way. The initial three nodes in $G_0$ are
labeled as 1, 2, and 3, respectively. In each new generation, only
the new nodes created at this generation are labeled, while the
labels of all old nodes remain unchanged, i.e., we label new nodes
as $M+1, M+2,\ldots, M+\Delta M$, where $M$ is the total number of
the pre-existing nodes and $\Delta M$ is the number of newly-created
nodes. Eventually, every node is labeled by a unique integer, at
time $n$ all nodes are labeled from 1 to $V_n=\frac{3^{n+1}+3}{2}$,
see Fig.~\ref{network}.

We locate the trap at node 1~\cite{Note1}, denoted as $i_T$. Note
that the particular selection we made for the trap location makes
the analytical computation process (that will be shown in detail in
the next section) easily iterated as we can identify the trap node
$i_T$ since the first generation. At each time step (taken to be
unity), the particle, starting from any node except the trap $i_T$,
moves to any of its nearest neighbors with equal probability. It is
easily seen that in the presence of the trap $i_T$ fixed on node 1,
the walker (particle) will be inevitably absorbed~\cite{Bobe05}.
This random walk process can be described by a specifying the set of
transition probabilities $W_{ij}$ for the particle of going from
node $i$ to node $j$. In fact, all the entries constitute a matrix
$\textbf{W}$ that is a $(V_n-1)$-order sub-matrix of
$(\textbf{Z}^{-1})\textbf{A}$ with the first row and column
corresponding to trap being removed, where $\textbf{Z}$ and
$\textbf{A}$ are separately the adjacency matrix and diagonal degree
matrix of $G_n$~\cite{Note2}.

Let $T_i$ be survival time for a walker initially placed at node $i$
to first reach the trap $i_T$. Then, the set of this interesting
quantity obeys the following recurrence equation~\cite{KeSn76}
\begin{equation}\label{MFPT1}
 T_i=\sum_j W_{ij}\,T_j+1,
\end{equation}
where $i\neq i_T$. Eq.~(\ref{MFPT1}) expresses the Markovian
property of the random walk, it may be recast in matrix notation as
\begin{equation}\label{MFPT2}
 \textbf{T}=\textbf{W}\,\textbf{T}+\textbf{e},
\end{equation}
where $\textbf{T}=(T_2,T_3,\cdots,T_{V_n})^\top$ (the superscript
$\top$ of the vector represents transpose) is a
($V_n-1$)-dimensional vector, $\textbf{e}$ is the
($V_n-1$)-dimensional unit vector $(1,1,\cdots,1)^\top$, and
$\textbf{W}$ is the transition matrix. From Eq.~(\ref{MFPT2}), we
can easily obtain
\begin{equation}\label{MFPT3}
 \textbf{T}=\textbf{L}\,\textbf{e},
\end{equation}
where
\begin{equation}\label{MFPT4}
 \textbf{L}=(\textbf{I}-\textbf{W})^{-1},
\end{equation}
in which $\textbf{I}$ is an identity matrix with an order
$(V_n-1)\times (V_n-1)$. Equation~(\ref{MFPT4}) is the fundamental
matrix of the Markov chain representing the unbiased random walk.
Actually, the matrix $\textbf{I}-\textbf{W}$ in Eq.~(\ref{MFPT4}) is
the normalized discrete Laplacian matrix of $G_n$ whose first row
and column that correspond to the trap node have been deleted.

Then, the mean first-passage time (MFPT), or the average of the mean
time to absorption, $\langle T \rangle_n$, which is the average of
$T_i$ over all initial nodes distributed uniformly over nodes in
$G_n$ other than the trap, is given by
\begin{equation}\label{MFPT5}
 \langle T
\rangle_n=\frac{1}{V_n-1}\sum_{i=2}^{V_n}
T_i=\frac{1}{V_n-1}\sum_{i=2}^{V_n}\sum_{j=2}^{V_n}{L_{ij}}.
\end{equation}

Equation~(\ref{MFPT5}) can be easily explained from the Markov chain
representing the random walk. In fact, the entry $L_{ij}$ of the
fundamental matrix $\textbf{L}$ for the Markov process denotes the
expected number of times that the process is in the transient state
$j$, being started in the transient state $i$.

The quantity of MFPT is very important since it measures the
efficiency of the trapping process: the smaller the MFPT, the higher
the efficiency, and vice versa. Equation~(\ref{MFPT5}) shows that
the problem of calculating $\langle T \rangle_n$ is reduced to
finding the sum of all elements of matrix $\textbf{L}$. Notice that
the order of $\textbf{L}$ is $(V_n-1)\times (V_n-1)$, where $V_n$
increases exponentially with $n$, as shown in Eq.~(\ref{Nt}). So,
for large $n$, it becomes difficult to obtain $\langle T \rangle_n$
through direct calculation from Eq.~(\ref{MFPT5}), one can compute
directly the MFPT only for the first several generations, see
Table~\ref{tab:AMTA1}. However, the recursive construction of PSFW
allows one to compute analytically MFPT to achieve a closed-form
solution, the derivation details of which will be given in next
section.

\begin{table}
\caption{The average of the mean time to absorption obtained by
direct calculation from Eq.~(\ref{MFPT5}). Since for large networks,
the computation of the MFPT from Eq.~(\ref{MFPT5}) is prohibitively
time and memory consuming, we calculate the MFPT for only the first
several generations.} \label{tab:AMTA1}
\begin{center}
\begin{tabular}{ccccc}
\hline \hline
\quad\quad $n$\quad\quad\quad & \quad\quad\  $V_n$ \quad\quad\quad  &\quad\quad\quad $\langle T\rangle_n$ \quad\quad\quad \\
\hline
0  & 3 & 4/2  \\
1  & 6 & 19/5  \\
2  & 15 & 101/14  \\
3 & 42 & 571/41 \\
4  & 123 & 3329/122  \\
5  & 366 & 19699/365  \\
6  & 1095 & 117401/1094  \\
7  & 3282 & 702091/3281  \\
8  & 9843 & 4205729/9842  \\
\hline \hline
\end{tabular}
\end{center}
\end{table}

\section{Exact solution for mean first-passage time}

Before deriving the general formula for MFPT, $\langle T \rangle_n$,
we first establish the scaling relation dominating the evolution of
$T_i^n$ with generation $n$, where $T_i^n$ is the trapping time for
a walk originating at node $i$ on the $n$th generation of PSFW.

\subsection{Evolution scaling for trapping time}

We begin by recording the numerical values of $T_i^n$. Obviously,
for all $n \geq 0$, $T_1^n=0$; for $n = 0$, it is a trivial case, we
have $T_2^0=T_3^0=2$. For $n \geq 1$, the values of $T_i^n$ can be
obtained straightforwardly via Eq.~(\ref{MFPT5}). In the generation
$n =1$, by symmetry we have $T_2^1=T_3^1=4$, $T_4^1=T_5^1=3$, and
$T_6^1=5$. Analogously, for $n =2$, the solutions are
$T_2^2=T_3^2=8$, $T_4^2=T_5^2=6$, $T_6^2=10$, $T_7^2=T_8^2=4$,
$T_9^2=T_{12}^2=8$, $T_{10}^2=T_{11}^2=10$, $T_{13}^2=T_{14}^2=5$,
and $T_{15}^2=9$. Table \ref{tab:AMTA2} lists the numerical values
of $T_i^n$ for some nodes up to $n=6$.

\begin{table}
\caption{Mean time to absorption $T_i^n$ for a random walker
starting from node $i$ on PSFW for various $n$. Notice that owing to
the obvious symmetry, nodes in a parenthesis are equivalent, since
they have the same trapping time. All the values are calculated
straightforwardly from Eq.~(\ref{MFPT5}).} \label{tab:AMTA2}
\begin{center}
\begin{tabular}{l|cccccccccc}
\hline \hline  $n\backslash i$  & (2,3) & (4,5) &  (6)\quad &(7,8) &(9,12) &(10,11) &(13,14) & (15) \\
\hline
0&2 \\
1&4&3&5\\
2&8&6&10&4&8&10&5&9\\
3&16&12&20&8&16&20&10&18\\
4&32&24&40&16&32&40&20&36\\
5&64&48&80&32&64&80&40&72\\
6&128&96&160&64&128&160&80&144\\
\hline \hline
\end{tabular}
\end{center}
\end{table}

The numerical values listed in Table \ref{tab:AMTA2} show that for a
given node $i$ we have $T_i^{n+1}=2\,T_i^n$. That is to say, upon
growth of PSFW from $n$ to generation $n+1$, the mean time to first
reach the trap doubles. For example,
$T_2^6=2\,T_2^5=4\,T_2^4=8\,T_2^3=16\,T_2^2=32\,T_2^1=64\,T_2^0=128$,
$T_4^6=2\,T_4^5=4\,T_4^4=8\,T_4^3=16\,T_4^2=32\,T_4^1=96$,
$T_7^6=2\,T_7^5=4\,T_7^4=8\,T_7^3=16\,T_7^2=64$, and so on. This is
a basic character of random walks on the PSFW, which can be
established from the arguments below.

Consider an arbitrary node $i$ in the PSFW $G_n$ after $n$
generation evolution of the network. From Eq.~(\ref{ki}), we know
that upon growth of PSFW to generation $n+1$, the degree $k_i$ of
node $i$ doubles. Let the mean transmit time for going from node $i$
to any of its $k_i$ old neighbors be $Y$; and let the mean transmit
time for going from any of its $k_i$ new neighbors to one of the
$k_i$ old neighbors be $Z$. Then we can establish the following
underlying backward equations
\begin{eqnarray}\label{MFPT6}
\left\{
\begin{array}{ccc}
Y&=&\frac{1}{2}+\frac{1}{2}(1+Z),\\
Z&=&\frac{1}{2}+\frac{1}{2}(1+Y),\\
 \end{array}
 \right.
\end{eqnarray}
which leads to $Y=2$. That is to say, the passage time from any node
$i$ ($i \in G_{n}$) to any node $j$ ($j\in G_{n}$) increases by a
factor of 2, upon the network growth from generation $n$ to
generation $n+1$. Thus, we have $T_i^{n+1}=2\,T_i^n$, which will be
useful for deriving the formula for the mean first-passage time in
the following text.

\subsection{Formula for the mean first-passage time}

Having obtained the scaling of mean transmit time for old nodes, we
now determine the average of the mean time to absorption, aiming to
derive an exact solution. We represent the set of nodes in $G_{n}$
as $\Lambda_n$, and denote the set of nodes created at generation
$n$ by $\overline{\Lambda}_n$. Thus we have
$\Lambda_n=\overline{\Lambda}_n\cup\Lambda_{n-1}$. For the
convenience of computation, we define the following quantities for
$m \leq n$:
\begin{equation}\label{MFPT7}
 T_{m,\text{total}}^n=\sum_{i\in \Lambda_m} T_i^n,
\end{equation}
and
\begin{equation}\label{MFPT8}
 \overline{T}_{m,\text{total}}^n=\sum_{i\in \overline{\Lambda}_m}
 T_i^n.
\end{equation}
Then, we have
\begin{equation}\label{MFPT9}
 T_{n,\text{total}}^n=T_{n-1,\text{total}}^n+\overline{T}_{n,\text{total}}^n.
\end{equation}
Next we will explicitly determine the quantity
$T_{n,\text{total}}^n$. To this end, we should firstly determine
$\overline{T}_{n,\text{total}}^n$.

We examine the mean time to absorption for the first several
generations of PSFW. In the case of $n=1$, by construction of the
PSFW, it follows that
$T^{1}_4=\frac{1}{2}(1+T^{1}_1)+\frac{1}{2}(1+T^{1}_2)$,
$T^{1}_5=\frac{1}{2}(1+T^{1}_1)+\frac{1}{2}(1+T^{1}_3)$, and
$T^{1}_6=\frac{1}{2}(1+T^{1}_2)+\frac{1}{2}(1+T^{1}_3)$. Thus,
\begin{eqnarray}\label{MFPT10}
\overline{T}_{1,\text{total}}^1&=&\sum_{i\in
\overline{\Lambda}_1}T^{1}_i=T^{1}_4+T^{1}_5+T^{1}_6\nonumber\\
&=&3+(T^{1}_1+ T^{1}_2+
T^{1}_3)=3+\overline{T}_{0,\text{total}}^1\,.
\end{eqnarray}
Similarly, for $n=2$ case,
$T^{2}_7=\frac{1}{2}(1+T^{2}_1)+\frac{1}{2}(1+T^{2}_4)$,
$T^{2}_8=\frac{1}{2}(1+T^{2}_1)+\frac{1}{2}(1+T^{2}_5)$,
$T^{2}_9=\frac{1}{2}(1+T^{2}_3)+\frac{1}{2}(1+T^{2}_5)$,
$T^{2}_{10}=\frac{1}{2}(1+T^{2}_3)+\frac{1}{2}(1+T^{2}_6)$,
$T^{2}_{11}=\frac{1}{2}(1+T^{2}_2)+\frac{1}{2}(1+T^{2}_6)$,
$T^{2}_{12}=\frac{1}{2}(1+T^{2}_2)+\frac{1}{2}(1+T^{2}_4)$,
$T^{2}_{13}=\frac{1}{2}(1+T^{2}_1)+\frac{1}{2}(1+T^{2}_2)$,
$T^{2}_{14}=\frac{1}{2}(1+T^{2}_1)+\frac{1}{2}(1+T^{2}_3)$, and
$T^{2}_{15}=\frac{1}{2}(1+T^{2}_2)+\frac{1}{2}(1+T^{2}_3)$, so that
\begin{eqnarray}\label{MFPT11}
\overline{T}_{2,\text{total}}^2&=&\sum_{i\in
\overline{\Lambda}_2}T^{2}_i=\sum_{i=7}^{15}T^{2}_i\nonumber\\
&=&3^2+2\,(T^{2}_1+ T^{2}_2+ T^{2}_3)+(T^{2}_4+ T^{2}_5+
T^{2}_6)\nonumber\\
&=&3^2+2\,\overline{T}_{0,\text{total}}^2+\overline{T}_{1,\text{total}}^2\,.
\end{eqnarray}
Proceeding analogously, it is not difficult to derive that
\begin{eqnarray}\label{MFPT12}
\overline{T}_{n,\text{total}}^n=3^n&+&\overline{T}^{n}_{n-1,\text{total}}+2\,\overline{T}^{n}_{n-2,\text{total}}+\dots\nonumber\\
&+&2^{n-2}\,\overline{T}^{n}_{1,\text{total}}+2^{n-1}\,\overline{T}^{n}_{0,\text{total}},
\end{eqnarray}
and
\begin{eqnarray}\label{MFPT13}
\overline{T}_{n+1,\text{total}}^{n+1}=3^{n+1}&+&\overline{T}^{n+1}_{n,\text{total}}+2\,\overline{T}^{n+1}_{n-1,\text{total}}+\dots\nonumber\\
&+&2^{n-1}\,\overline{T}^{n+1}_{1,\text{total}}+2^{n}\,\overline{T}^{n+1}_{0,\text{total}}\,,
\end{eqnarray}
where $3^n$ and $3^{n+1}$ are indeed the numbers of nodes generated
at generations $n$ and $n+1$, respectively. Equation~(\ref{MFPT13})
minus Eq.~(\ref{MFPT12}) times 4 and making use of the relation
$T_i^{n+1}=2\,T_i^n$, one gets
\begin{equation}\label{MFPT14}
\overline{T}^{n+1}_{n+1,\text{total}}-3^{n+1}=\overline{T}^{n+1}_{n,\text{total}}+4\,\big(\overline{T}^{n}_{n,\text{total}}-3^n\big),
\end{equation}
which may be rewritten as
\begin{equation}\label{MFPT15}
\overline{T}^{n+1}_{n+1,\text{total}}=6\,\overline{T}^{n}_{n,\text{total}}-3^n.
\end{equation}
Using $\overline{T}_{1,\text{total}}^1=11$, Eq.~(\ref{MFPT15}) is
solved inductively
\begin{equation}\label{MFPT16}
\overline{T}^{n}_{n,\text{total}}=\frac{5}{3}\times 6^n+3^{n-1}\,.
\end{equation}
Substituting Eq.~(\ref{MFPT16}) for
$\overline{T}^{n}_{n,\text{total}}$ into Eq.~(\ref{MFPT9}), we have
\begin{eqnarray}\label{MFPT17}
T_{n,\text{total}}^n&=&T_{n-1,\text{total}}^n+\frac{5}{3}\times 6^n+3^{n-1}\nonumber\\
&=&2\,T_{n-1,\text{total}}^{n-1}+\frac{5}{3}\times 6^n+3^{n-1}\,.
\end{eqnarray}
Considering the initial condition $T_{0,\text{total}}^0=4$,
Eq.~(\ref{MFPT17}) is resolved by induction to yield
\begin{equation}\label{MFPT18}
T_{n,\text{total}}^n=\frac{5}{2} \times 6^n+3^n+2^{n-1} \,.
\end{equation}
Plugging the last expression into Eq.~(\ref{MFPT5}), we arrive at
the accurate formula for the average of the mean time to absorption
at the trap located at node 1 on the $n$th of the pseudofractal
scale-free web:
\begin{eqnarray}\label{MFPT19}
 \langle T
\rangle_n &=&\frac{1}{V_n-1}\sum_{i=2}^{V_n}
T_i=\frac{1}{V_n-1}T_{n,\text{total}}^n \nonumber\\
&=&\frac{5\times 6^n+2\times 3^n+2^n}{3^{n+1}+1}\,.
\end{eqnarray}
We have checked our analytic formula against numerical values quoted
in Table~\ref{tab:AMTA1}. For the range of $0 \leq n \leq 8$, the
values obtained from Eq.~(\ref{MFPT19}) completely agree with those
numerical results on the basis of the direct calculation through
Eq.~(\ref{MFPT5}). This agreement serves as an independent test of
our theoretical formula.

We continue to show how to represent MFPT as a function of network
order, with the aim of obtaining the scaling between these two
quantities. Recalling Eq.~(\ref{Nt}), we have $3^{n+1}=2V_n-3$ and
$n+1=\log_3(2V_n-3)$. Hence, Eq.~(\ref{MFPT19}) can be recast as
\begin{equation}\label{MFPT20}
 \langle T
\rangle_n
=\frac{\frac{5}{6}(2V_n-3)^{1+\frac{\ln2}{\ln3}}+\frac{2}{3}(2V_n-3)+\frac{1}{2}(2V_n-3)^{\frac{\ln2}{\ln3}}}{2V_n-2}.
\end{equation}
For large network, i.e., $V_n\rightarrow \infty$,
\begin{equation}\label{MFPT21}
\langle T \rangle_n \sim (V_n)^{\frac{\ln2}{\ln3}},
\end{equation}
where the exponent $\frac{\ln2}{\ln 3}<1$. Thus, in the large limit
of network order $V_n$, the MFPT increases algebraically with
increasing order of the network.


The above scaling of the MFPT with network order is different from
those previously obtained for other media. For example, on regular
lattices with large order $N$, the leading behavior of MFPT $\langle
T \rangle$ is $\langle T \rangle \sim N^2$, $\langle T \rangle \sim
N\ln N$, and $\langle T \rangle \sim N$ for dimensions $d=1$, $d=2$,
and $d=3$, respectively~\cite{Mo69}. Again for instance, on planar
Sierpinski gasket~\cite{KaBa02PRE} and Sierpinski
tower~\cite{KaBa02IJBC} in 3 Euclidean dimensions, the asymptotic
behavior scales separately as $\langle T \rangle \sim N^{1.464}$ and
$\langle T \rangle \sim N^{1.293}$. At last, on T-fractal, the
leading asymptotic scaling behaves as $\langle T \rangle \sim
N^{1.631}$~\cite{Ag08}. Thus, in contrast to regular lattices,
Sierpinski fractals, and T-fractal, the trapping process on the
pseudofractal scale-free web is more efficient. It is expected that
the efficiency of trapping process on stochastic scale-free networks
is also high, since they have similar structural features as the
PSFW.

Why is the MFPT for the PFSW far smaller than that for other
lattices? We speculate that the heterogeneity of the pseudofractal
scale-free web is responsible for this distinction, which may be
seen from the following heuristic argument. In the PFSW, there are a
few nodes with large degree that are connected to most nodes in the
web, which results in the logarithmic scaling of the average path
length with network order~\cite{DoGoMe02,ZhZhCh07}. A walker
starting from some node will arrive at `large' nodes with ease.
Since `large' nodes, including the trap node, are linked to one
another, so the walker can find the trap in a short time. While for
other regular lattices, they are almost homogeneous, and their
average path lengths are much larger than that of the PFSW. This
leads to a long MFPT. It should be noted that we only give a
possible explanation for the shorter MFPT on the PFSW, the genuine
reason for this deserves further study in the future.

\section{Conclusions}

In summary, we have investigated the classic trapping problem on a
deterministically growing network, named pseudofractal scale-free
web (PSFW) that can reproduce some remarkable properties of various
real-life networks, such as scale-free feature and small-world
behavior, and thus can mimic some real systems to some extent (to
what extent it does is still an open question). With the help of
recursion relations derived from the structure of PSFW, we have
obtained the solution for the MFPT for random walks on PSFW, with a
trap fixed at a hub node. The exact result shows that the MFPT
increases as a power-law function of network order with the exponent
less than 1, which is in contrast to the well-known previously
obtained results that for regular lattices, Sierpinski fractals, and
T-graph with order $N$, their MFPT $ \langle T \rangle $ behaves as
$ \langle T \rangle \sim N^{\alpha}$ with $\alpha >1$. Therefore,
the structure of PSFW has a profound impact on the trapping problem
on it. To the best of our knowledge, our result may be the first
exact scaling about MFPT for random walks on scale-free small-world
networks.

We should stress that although we have only computed the MFPT for a
particular deterministic network, our analytical technique could
guide and shed light on related studies for other deterministic
networks by providing a paradigm for calculating the MFPT. Moreover,
since exact solutions can serve for a guide to and a test of
approximate solutions or numerical simulations, we also believe our
accurate closed-form solutions can prompt related studies on
stochastic networks. At last, as future work, it is interesting to
compute higher moments of trapping time for the PSFW and compare the
scaling with that of homogeneous fractal lattices~\cite{HaRo08}.
Another future job of interest is to study the case when the trap is
mobile instead of being fixed at one of the hub nodes of the web.

\subsection*{Acknowledgment}

We would like to thank Yichao Zhang for preparing this manuscript.
This research was supported by the National Basic Research Program
of China under grant No. 2007CB310806, the National Natural Science
Foundation of China under Grant Nos. 60704044, 60873040 and
60873070, Shanghai Leading Academic Discipline Project No. B114, and
the Program for New Century Excellent Talents in University of China
(NCET-06-0376).

\end{document}